\documentclass[nofootinbib,onecolumn,preprintnumbers,amsmath,amssymb]{revtex4}
\def\bea{\begin{eqnarray}}
\def\eea{\end{eqnarray}}
\newcommand{\be}{\begin{equation}}
\newcommand{\ee}{\end{equation}}

\newcommand{\bn}{\begin{eqnarray}}
\newcommand{\en}{\end{eqnarray}}

\def\bea{\begin{eqnarray}}
\def\eea{\end{eqnarray}}

\usepackage{amsmath}
\usepackage{amssymb}
\usepackage{graphicx}
\usepackage{bm}
\usepackage{epstopdf}
\usepackage{slashed}
\usepackage{appendix}

\usepackage{xcolor}
\usepackage[unicode=true,
 bookmarks=true,bookmarksnumbered=false,bookmarksopen=false,
 breaklinks=false,pdfborder={0 0 1},backref=false,colorlinks=false]
 {hyperref}

\begin{document}

\title{External Sources in Pseudo-Quantum Electrodynamics}

\author{L.H.C. Borges}
\email{luizhenrique.borges@ufla.br}

\affiliation{Departamento de Física, Universidade Federal de Lavras,
Caixa Postal 3037, 37200-900 Lavras-MG, Brazil.
}

\author{A.A. Nogueira}
\email{andsogueira@hotmail.com}

\affiliation{Santa Catarina State University, Department of Physics, Joinville, SC, 89219-710, Brazil.}

\author{F.A. Barone}
\email{fbarone@unifei.edu.br}

\affiliation{IFQ - Universidade Federal de Itajub\'a, Av. BPS 1303, Pinheirinho, Caixa Postal 50, 37500-903, Itajub\'a, MG, Brazil.}

\begin{abstract}
In this letter, we investigate the interaction between stationary point-like external sources within the framework of Pseudo-Quantum Electrodynamics (PQED). Besides ordinary electric charges, we consider configurations involving Dirac points and generalized topological sources. For the configurations analyzed, we compute the corresponding interaction energies, forces, torques, and electromagnetic field configurations. Furthermore, we compare the results obtained in PQED with those arising in ordinary planar Maxwell electrodynamics, showing that the nonlocal character of PQED gives rise to physical effects with no counterpart in the conventional local planar theory. In particular, our results show that topological sources induce intrinsically anisotropic and non-central interactions, leading to the emergence of nontrivial torques between the interacting objects. The directional character of these interactions suggests possible applications in the effective description of topological defects and anisotropic phenomena in planar condensed matter systems.
\end{abstract}

\maketitle


Pseudo-Quantum Electrodynamics (PQED), which is a nonlocal theory, has emerged as an important effective field-theoretical framework for the description of planar systems in which charged matter is confined to two spatial dimensions while the electromagnetic interaction remains intrinsically three-dimensional \cite{Marino1993,Marino1993b}. In contrast to conventional Quantum Electrodynamics in $(2+1)$ dimensions, PQED correctly reproduces the $1/r$ Coulomb interaction between static charges confined to the plane, thereby preserving essential features of the electromagnetic interaction observed in realistic condensed matter systems. This property makes PQED particularly suitable for the investigation of two-dimensional Dirac materials, especially graphene, where the electronic quasiparticles are effectively restricted to a planar geometry while the gauge field propagates in the surrounding three-dimensional space. Consequently, PQED has been widely employed in the study of several electronic properties of planar condensed matter systems 
\cite{Alves2013,Teber2014,Barnes2014,Nascimento2015,Marino2015,Menezes2016,Kotikov2016,Alves2017,Menezes2017,Magalhaes2020,Olivares2020,Ozela2022,Albino2022,Olivares2022,Kotov2012}.
Extensions of PQED have been investigated in the framework of Lee-Wick, Chern--Simons, and Proca gauge theories 
\cite{Alves2019,Ozela2019,Ozela2023,Alves2018,Xing2023,Neves2025a,Neves2025b,Neves2026}. 

Substrates, impurities, defects, and externally applied fields may significantly influence the electromagnetic environment surrounding planar materials \cite{Kotov2012, Katsnelson2012}. Within a field-theoretical description, such effects can be incorporated through classical external sources coupled to the gauge sector. In the context of PQED, the role of external sources has been studied primarily for point-like electric charges. Nevertheless, since PQED provides a more general framework for describing electromagnetic interactions in planar media, it is natural to investigate how the theory responds to broader classes of external sources and whether qualitatively new phenomena may arise from them.

This letter is devoted to the investigation of physical effects produced by stationary point-like external sources in the framework of PQED. In particular, we study the interaction between distinct pairs of external sources coupled to the gauge sector of the theory. 
Besides the usual point-like electric charges, we focus on configurations involving two classes of nontrivial localized sources, namely Dirac points and topological sources, which were originally proposed in Ref. \cite{LHCBMCS} in the context of Chern-Simons electrodynamics and subsequently investigated in the framework of other local planar field theories \cite{LHCBHMCS,LHCBLVMCS,OliveiraLemos2025}.

For all configurations considered throughout this work, we show that the presence of topological sources induces nontrivial torques between the interacting objects, revealing the anisotropic nature of the corresponding interactions. We also determine the electromagnetic field configurations generated by the point-like sources analyzed in this letter and compare the results obtained in PQED with those arising in ordinary planar Maxwell electrodynamics \cite{LHCBMCS}. This comparison demonstrates that the nonlocal character of PQED gives rise to physical effects associated with Dirac points that have no counterpart in the conventional planar Maxwell theory. More generally, the introduction of topological sources into the framework of PQED has consequences that extend well beyond modifying the power laws governing the interactions between field sources so as to reproduce the corresponding behavior of Maxwell electrodynamics in $(3+1)$ dimensions. Rather, it reveals a richer interaction structure associated with generalized topological sources and predicts genuinely new physical phenomena.

Our results indicate that topological interactions in PQED may lead to nontrivial anisotropic effects, which could be relevant for the effective description of planar anisotropic systems and topological defects in two-dimensional condensed matter systems.

Along the letter we consider a Minkowski $(2+1)$-dimensional spacetime with metric $\eta^{\mu\nu}=(1,-1,-1)$. The Levi-Civita tensor is denoted by $\epsilon^{\mu\nu\lambda}$ with $\epsilon^{012}=1$.



The gauge sector of PQED is described by the following Lagrangian density \cite{Entropy}:
\begin{eqnarray}
\label{Lagrangian}
{\cal L}= -\frac{1}{2}F_{\mu\nu}\frac{1}{\Box^{1/2}}F^{\mu\nu}-\frac{\xi}{2}A_{\mu}\frac{\partial^{\mu}\partial^{\nu}}{\Box^{1/2}}A_{\nu}-J^{\mu}A_{\mu} \ ,
\end{eqnarray}
 where $A^{\mu}$ denotes the electromagnetic field, $F^{\mu\nu}=\partial^{\mu}A^{\nu}-\partial^{\nu}A^{\mu}$ represents the field strength tensor, $J^{\mu}$ corresponds to the external current, $\xi$ is the gauge-fixing parameter, and $\Box=\partial^{\mu}\partial_{\mu}$ is the d'Alembert operator.

We note that the model described by Eq. (\ref{Lagrangian}) is nonlocal, while still preserving causality \cite{Amaral1992}, satisfying the Huygens principle, maintaining unitarity \cite{Marino2014}, and exhibiting both scale invariance \cite{Dudal2019,Heydeman2020} and gauge invariance \cite{Marino1993}.

The model (\ref{Lagrangian}) can be rewritten as
\begin{eqnarray}
\label{LagrnagianOP}
{\cal L}\rightarrow \frac{1}{2}A^{\mu}{\cal{O}}_{\mu\nu}A^{\nu}-J^{\mu}A_{\mu} \ ,
\end{eqnarray}
where the differential operator is defined as
\begin{eqnarray}
 {\cal{O}}_{\mu\nu}=  \frac{2\Box\eta_{\mu\nu}-\left(2+\xi\right)\partial_{\mu}\partial_{\nu}}{\Box^{1/2}} \ .
\end{eqnarray}

The propagator $D^{\mu\nu}\left(x,y\right)$  is the inverse of the operator ${\cal{O}}^{\mu\nu}$, as follows
\begin{eqnarray}
  {\cal{O}}^{\mu\nu}D_{\nu\lambda}=\eta_{\ \lambda}^{\mu}\delta^{3}\left(x-y\right) \ . 
\end{eqnarray}

By applying standard techniques from field theory, one finds that the propagator takes the form
\begin{eqnarray}
\label{propagatorrr}
D^{\mu\nu}\left(x,y\right)=\int\frac{d^{3}p}{(2\pi)^{3}}\left[\frac{1}{2\left(-p^{2}\right)^{1/2}}\left(\eta^{\mu\nu}-\frac{p^{\mu}p^{\nu}}{p^{2}}\right)-\frac{1}{\xi}\frac{1}{\left(-p^{2}\right)^{1/2}}\frac{p^{\mu}p^{\nu}}{p^{2}}\right]e^{-ip\cdot(x-y)} \ .
\end{eqnarray}

Throughout this letter, we shall adopt the Landau gauge $(\xi \rightarrow \infty)$ \cite{Magalhaes2020}. Under this choice, the propagator in Eq. (\ref{propagatorrr}) reduces to
\begin{eqnarray}
\label{propagator}
D^{\mu\nu}\left(x,y\right)=\int\frac{d^{3}p}{(2\pi)^{3}}\frac{1}{2\left(-p^{2}\right)^{1/2}}\left(\eta^{\mu\nu}-\frac{p^{\mu}p^{\nu}}{p^{2}}\right)e^{-ip\cdot(x-y)} \ .
\end{eqnarray}

As discussed in Refs. \cite{LHCBMCS,LHCBHMCS,LHCBLVMCS,LHCBFABLW}, since the theory considered here is quadratic, the contribution of stationary field sources to the ground-state energy of the system can be obtained from the expression
\begin{equation}
\label{zxc1}
E=\frac{1}{2T}\int\int d^{3}x\ d^{3}y J^{\mu}(x)D_{\mu\nu}(x,y)J^{\nu}(y)\ ,
\end{equation}
where $T$ is the time variable and it is implicit the limit $T\rightarrow\infty$.

In the first case, the field sources are described by the external current
\begin{eqnarray}
\label{corre1Em}
J^{CC}_{\mu}({\bf x})=q_{1}\eta_{\ \mu}^{0}\delta^{2}\left({\bf x}-{\bf a}_ {1}\right)+q_{2}\eta_{\ \mu}^{0}\delta^{2}\left({\bf x}-{\bf a}_ {2}\right) \ ,
\end{eqnarray}
where two spatial Dirac delta functions are localized at the positions ${\bf a}_{1}$ and ${\bf a}_{2}$. The parameters $q_{1}$ and $q_{2}$ denote the coupling strengths between the field and the delta functions, and they can be interpreted as electric charges. The superscript $CC$ indicates that the configuration corresponds to the interaction between two point-like charges.

By substituting Eqs.~(\ref{propagator}) and (\ref{corre1Em}) into Eq.~(\ref{zxc1}), neglecting the self-interaction terms (i.e., the interaction of each point charge with itself), and carrying out the integrations in the order $d^{2}{\bf x}$, $d^{2}{\bf y}$, $dx^{0}$, $dp^{0}$, and $dy^{0}$, while employing the Fourier representation of the Dirac delta function, $\delta(p^{0})=\int dx^{0}/(2\pi)\,\exp(-ip^{0}x^{0})$, and identifying the time interval as $T=\int dy^{0}$, we arrive at
\begin{eqnarray}
\label{Ener2EM}
E^{CC}=\frac{q_{1}q_{2}}{2}\int\frac{d^{2}{\bf p}}{(2\pi)^{2}}\frac{e^{i{\bf p}\cdot{\bf a}}}{\left({\bf p}^2\right)^{1/2}} \ .
\end{eqnarray}

Using the fact that \cite{Gradshteyn}
\begin{eqnarray}
\label{int4EM}
\int\frac{d^{2}{\bf p}}{(2\pi)^{2}}\frac{e^{i{\bf p}\cdot{\bf a}}}{\left({\bf p}^2\right)^{1/2}}=\frac{1}{2\pi a} \ ,
\end{eqnarray}
we obtain
\begin{eqnarray}
\label{Ener3EM}
E^{CC}=\frac{q_{1}q_{2}}{4\pi a} \ ,
\end{eqnarray}
where we have introduced ${\bf a} = {\bf a}_{1} - {\bf a}_{2}$, representing the separation vector between the two electric charges, with $a=\mid\bf{a}\mid$. 
From this point onward, we shall use this notation to denote the separation distance between the external sources.

Eq. (\ref{Ener3EM}) stands for the interaction energy between two point-like charges in the framework of  PQED and reproduces the Coulomb potential, as in Maxwell electrodynamics in $(3+1)$ dimensions \cite{FABGH2008}. This result is already well known and constitutes one of the central and well-established features of PQED \cite{Entropy}.

Let us now investigate whether other types of interactions may arise from non-trivial external sources. To this end, we consider a second type of point-like source in PQED, namely a topological source. We begin by analyzing a system composed of two such sources located at positions ${\bf a}_{1}$ and ${\bf a}_{2}$ \cite{LHCBMCS}
\begin{eqnarray}
\label{sourceet2}
J^{TT}_{\mu}\left(x\right)=\epsilon_{\mu}^{\ \alpha\beta}V_{\alpha}
\partial_{\beta}\delta^{2}\left({\bf x}-{\bf a}_{1}\right) 
+\epsilon_{\mu}^{\ \alpha\beta}U_{\alpha}\partial_{\beta}
\delta^{2}\left({\bf x}-{\bf a}_{2}\right)\ ,
\end{eqnarray}
where the superscript $TT$ indicates that the configuration corresponds to the interaction between two topological sources and 
$V^{\alpha}=\left(V^{0}, {\bf{V}}\right)$ and $U^{\alpha}=\left(U^{0}, {\bf{U}}\right)$ are two Minkowski $3$-pseudo-vectors taken to be constant and uniform in the reference frame where the calculations are performed.

By inserting the source (\ref{sourceet2}) and the propagator (\ref{propagator}) into Eq. (\ref{zxc1}), neglecting the self-interaction terms, and following the same procedure adopted previously, we obtain
\begin{eqnarray}
\label{EnerTEM}
E^{TT}&=&\frac{1}{2}\Biggl[\left({\bf V}\cdot {\bf U}-V^{0}U^{0}\right)\int\frac{d^{2}{\bf p}}{(2\pi)^{2}}{\left({\bf p}^2\right)^{1/2}e^{i{\bf p}\cdot{\bf a}}}\nonumber\\
&
&+\left({\bf{V}}
\cdot{\bf\nabla}_{{\bf a}}\right)\left({\bf{U}}\cdot{\bf\nabla}
_{{\bf a}}\right)\int\frac{d^{2}{\bf p}}{(2\pi)^{2}}\frac{e^{i{\bf p}\cdot{\bf a}}}{\left({\bf p}^2\right)^{1/2}}\
\Biggr] \ .
\end{eqnarray}
where we defined the differential operator
\begin{eqnarray}
\label{exchange}
{\bf\nabla}_{{\bf a}}=\left(\frac{\partial}{\partial a^{1}},\frac{\partial}{\partial a^{2}}\right) \ .
\end{eqnarray}

Substituting the result (\ref{int4EM}) in the energy (\ref{EnerTEM}), using the fact that \cite{Gradshteyn} 
\begin{eqnarray}
\label{int4EMMM}
\int\frac{d^{2}{\bf p}}{(2\pi)^{2}}{\left({\bf p}^2\right)^{1/2}e^{i{\bf p}\cdot{\bf a}}}=-\frac{1}{2\pi a^{3}} \ ,
\end{eqnarray}
and carrying out the calculations, we obtain
\begin{eqnarray}
\label{EnerTEM2}
E^{TT}=\frac{1}{4\pi a^{3}}\Biggl[V^{0}U^{0}-2\left({\bf{V}}\cdot{\bf{U}}\right)+3\frac{\left({\bf{V}}
\cdot{\bf{a}}\right)\left({\bf{U}}\cdot{\bf{a}}\right)}{{\bf {a}}^{2}} 
\Biggr] \ .
\end{eqnarray}
An interesting feature of the interaction energy in Eq. (\ref{EnerTEM2}) is its highly anisotropic nature. In contrast to ordinary central interactions, which depend exclusively on the distance between the sources, the present interaction explicitly depends on the relative orientation between the vectors ${\bf V}$, ${\bf U}$, and the separation vector ${\bf a}$.

It is worth emphasizing the physical interpretation of the Eq. (\ref{EnerTEM2}).
In the particular case where $U^{0}=V^{0}=0$, the interaction energy reduces to that between two point-like electric dipoles in PQED with dipole moments
${\bf d}_{(1)}=({\bf V}\times\hat{\bf z})$ and
${\bf d}_{(2)}=({\bf U}\times\hat{\bf z})$. This discussion is not trivial, and we present it in Appendix~\ref{A}. 
This result exactly reproduces the interaction energy between two point-like electric dipoles in the conventional Maxwell electrodynamics in $(3+1)$ dimensions.
A similar interpretation also arises in the usual planar Maxwell electrodynamics, where, for the purely spatial case, the topological source behaves as an effective electric dipole \cite{LHCBMCS} .

On the other hand, for the purely temporal case (${\bf U}={\bf V}={ 0}$), the interaction energy (\ref{EnerTEM2})  becomes that between two point-like magnetic dipoles with magnetic moments
$m_{1}=V^{0}$ and $m_{2}=U^{0}$
(see Appendix \ref{AppMagneticDipole}).
Remarkably, the resulting expression is identical to the interaction energy between two magnetic dipoles in the conventional Maxwell electrodynamics in $(3+1)$ dimensions whose magnetic moments are oriented perpendicular to the plane.
It is important to emphasize that this correspondence has no analogue in ordinary planar Maxwell electrodynamics, where  point-like magnetic dipoles do not interact (see Appendix \ref{AppMagneticDipole}). 

The interaction force between two topological field sources reads
\begin{eqnarray}
\label{fortptp} 
{\bf{F}}^{TT}&=&-{\bf\nabla}_{{\bf a}}E^{TT}\nonumber\\
&=&\frac{3}{4\pi a^{4}}\Biggl\{\left[V^{0}U^{0}-2\left({\bf{V}}\cdot{\bf{U}}\right)+5\frac{\left({\bf{V}}\cdot{\bf{a}}\right)\left({\bf{U}}\cdot{\bf{a}}\right)}{a^{2}}\right]{\hat{a}}\nonumber\\
&
&-\frac{\left({\bf{V}}\cdot{\bf{a}}\right)}{a}{\bf{U}}-\frac{\left({\bf{U}}\cdot{\bf{a}}\right)}{a}{\bf{V}}\Biggr\} \ ,
\end{eqnarray}

where  ${\hat{a}}$ is an unit vector pointing in the direction of the vector ${\bf {a}}$.

It is instructive to analyze the anisotropic character of the interaction force (\ref{fortptp}) through its corresponding field lines. For this purpose, we consider the particularly simple configuration defined by
\[
U^{0}=V^{0}=0,
\qquad
{\bf V}=V_{x}\hat{x},
\qquad
{\bf U}=U_{y}\hat{y}.
\]
\begin{figure}[h!]
    \centering
    \includegraphics[width=0.5\textwidth]{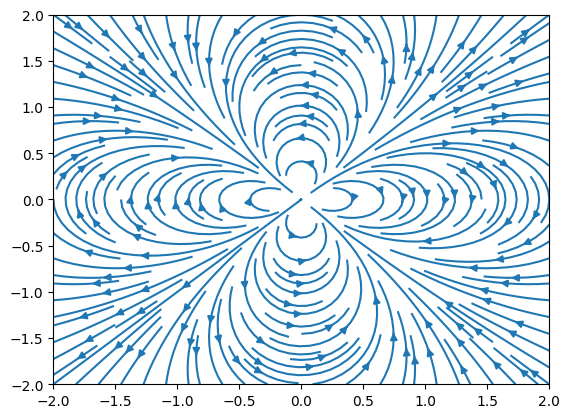}
    \caption{Force lines associated with the interaction between two topological sources for the particular configuration $U^{0}=V^{0}=0$, ${\bf V}=V_{x}\hat{x}$, and ${\bf U}=U_{y}\hat{y}$. The field lines were obtained from Eq. (\ref{fortptp}) after normalization by the factor $1/(V_{x}U_{y})$. The plot clearly exhibits the strongly anisotropic and non-central character of the interaction, revealing the existence of preferred directions in the plane. The horizontal and vertical axes correspond to the components $a_{x}$ and $a_{y}$, respectively.
}
    \label{fig1}
\end{figure}

In Fig. \ref{fig1}, we display the force lines obtained from Eq. (\ref{fortptp}), normalized by the factor $1/(V_{x}U_{y})$. The horizontal and vertical axes correspond to the components $a_{x}$ and $a_{y}$, respectively. The resulting pattern clearly reveals the strongly anisotropic and non-central nature of the interaction, since the force field exhibits a pronounced angular dependence and does not follow the radial structure characteristic of ordinary Coulomb-like interactions.

Due to the anisotropic character of the interaction energy (\ref{EnerTEM2}), it is natural to investigate whether the system exhibits physical effects beyond the interaction force (\ref{fortptp}). To this end, we consider a configuration in which the separation $a$ between the topological sources is kept fixed while the vector ${\bf a}$ is allowed to rotate in the plane. In this case, the interaction energy depends explicitly on the angular orientation of the system, allowing for the emergence of a nontrivial torque acting on the topological sources.

For simplicity, we assume $V^{0}=U^{0}=0$, ${\bf V}=V_{x}\hat{x}$, ${\bf U}=U_{y}\hat{y}$, and
\[
{\bf a}=a\left[\cos(\theta)\hat{x}+\sin(\theta)\hat{y}\right],
\]
where $\theta$ is the usual polar angle in the plane. Under these assumptions, Eq.~(\ref{EnerTEM2}) reduces to
\begin{equation}
E^{TT}(V^{0}=U^{0}=0,\theta)=\frac{3V_{x}U_{y}}{4\pi a^{3}}\cos(\theta)\sin(\theta)\ ,
\end{equation}
and we have a torque on the whole system 
\begin{eqnarray}
\label{TTT}
{\tau}^{TT}&=&-\frac{\partial}{\partial\theta}E^{TT}(V^{0}=U^{0}=0,\theta)
=-\frac{3V_{x}U_{y}}{4\pi a^{3}}\cos\left(2\theta\right)\ .
\end{eqnarray}

\begin{figure}[h!]
    \centering
    \includegraphics[width=0.45\textwidth]{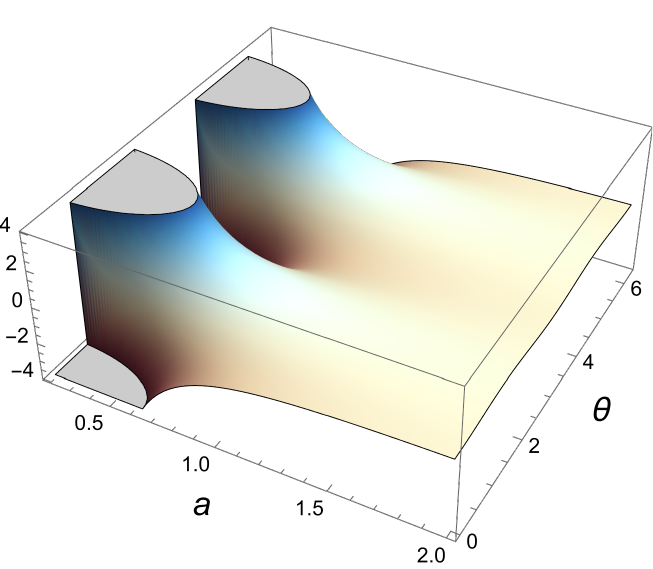}
    \caption{The torque given in Eq. (\ref{TTT}), multiplied by $\frac{4\pi}{3V_{x}U_{y}}$.}
    \label{fig2}
\end{figure}

We can observe this the torque vanishes for $\theta=\pi/4,\ 3\pi/4,\ 5\pi/4,\ 7\pi/4$, while its maximum magnitude is achieved for $\theta=0,\ \pi/2,\ \pi,\ 3\pi/2,\ 2\pi$. 
In Fig. \ref{fig2}, it is shown the general behavior of the torque (\ref{TTT})  as a function of $a$ and $\theta$.

We now consider another nontrivial external source, namely the Dirac point, defined by \cite{LHCBMCS,LHCBHMCS,LHCBLVMCS}
\begin{eqnarray}
\label{dpcharge}
J^{\mu}_{(D)}\left({\bf x}\right)=-2\pi i\Phi\int\frac{d^{3}p}{\left(2\pi\right)^{3}}
\delta\left(p^{0}\right)\epsilon^{0\mu\alpha}p_{\alpha}
e^{-i p\cdot x}e^{-i {\bf p}\cdot{\bf a}_{1}} \ ,
\end{eqnarray}
where $\Phi$ is a constant with dimension of magnetic flux, and ${\bf a}_{1}$ is a spatial vector specifying the position of the Dirac point. 

As discussed in Ref. \cite{LHCBMCS}, the topological source constitutes a generalization of the Dirac point. In particular, the Dirac point is recovered in the special limit where the topological source becomes purely temporal, namely
\[
V^{0}=-\Phi,
\qquad
{\bf V}=0,
\]
or equivalently,
\[
V^{\mu}=-\Phi\,\eta^{\mu0}.
\]
In this limit, the vectorial sector responsible for the anisotropic character of the topological source vanishes.

Therefore, by taking ${\bf U}={\bf V}=0$ in Eq. (\ref{EnerTEM2}), we recover the interaction energy between two Dirac points. In this configuration, the first Dirac point is located at ${\bf a}_{1}$ and carries magnetic flux $V^{0}=-\Phi_{1}$, while the second one is located at ${\bf a}_{2}$ with magnetic flux $U^{0}=-\Phi_{2}$. So that, the energy (\ref{EnerTEM2}) becomes
\begin{eqnarray}
\label{Ener5EM}
E^{TT}\left({\bf U}={\bf V}=0,V^{0}=-\Phi_{1}, U^{0}=-\Phi_{2} \right)=E^{DD}=\frac{\Phi_{1}\Phi_{2}}{4\pi a^{3}} \ ,
\end{eqnarray}
where the super-index $DD$ means that we
have a system composed by two Dirac points.

The interaction energy given in Eq. (\ref{Ener5EM}) is an effect that arises exclusively within the framework of PQED, since this interaction does not occur in standard Maxwell electrodynamics in $(2+1)$ dimensions \cite{LHCBMCS}.

From the Eq. (\ref{fortptp}), the interaction force between two Dirac points reads
\begin{eqnarray}
\label{For2EM}
{\bf{F}}^{TT}\left({\bf U}={\bf V}=0,V^{0}=-\Phi_{1}, U^{0}=-\Phi_{2} \right)={\bf{F}}^{DD}=\frac{3\Phi_{1}\Phi_{2}}{4\pi a^{4}}{\hat{a}}  \ .
\end{eqnarray}

As shown in Ref.~\cite{LHCBMCS}, in Chern--Simons electrodynamics, a Dirac point introduced as an external source behaves as a point-like electric charge. However, this result does not imply that the Dirac point considered here in PQED is phenomenologically equivalent to an ordinary point charge in Chern--Simons electrodynamics. Although Dirac points are introduced as external sources in both theories, the physical effects they produce are governed by the corresponding gauge dynamics. In Chern--Simons electrodynamics, the Chern--Simons term gives rise to the equivalence demonstrated in Ref.~\cite{LHCBMCS}. By contrast, the gauge sector of the PQED model considered here contains no Chern--Simons term, and therefore no such equivalence arises. It is worth emphasizing that the inclusion of a Chern--Simons or pseudo-Chern--Simons term in the gauge sector of PQED would define a different theory. Whether an external Dirac point would behave as a point-like electric charge in such a framework remains to be investigated. Consequently, this possible correspondence cannot be inferred directly from the result obtained in Ref.~\cite{LHCBMCS} and requires a separate analysis of PQED supplemented by a Chern--Simons or pseudo-Chern--Simons term \cite{BorgesBarone}.

In the following, we analyze the interaction between a point-like charge and a topological source, described by the external current \cite{LHCBMCS}
\begin{eqnarray}
\label{corre1Emmm}
J^{CT}_{\mu}({\bf x})=q\eta_{\ \mu}^{0}\delta^{2}\left({\bf x}-{\bf a}_ {1}\right)+
\epsilon_{\mu}^{\ \alpha\beta}V_{\alpha}\partial_{\beta}
\delta^{2}\left({\bf x}-{\bf a}_{2}\right) \ ,
\end{eqnarray}
where the point-like charge is located at ${\bf a}_{1}$, whereas the topological source is centered at ${\bf a}_{2}$. The superscript $CT$ denotes the interaction between a charge and a topological source.

In this configuration, the interaction energy takes the form
\begin{eqnarray}
\label{dpplc}
E^{CT}&=& \frac{q}{2}\left[\left({\bf{V}}\times{\hat{z}}\right)\cdot{\bf\nabla}_{{\bf a}}\right]\int\frac{d^{2}{\bf p}}{(2\pi)^{2}}\frac{e^{i{\bf p}\cdot{\bf a}}}{\left({\bf p}^2\right)^{1/2}}\nonumber\\
&=&-\frac{q}{4\pi a^{3}}\left[\left({\bf{V}}\times{\hat{z}}\right)\cdot{\bf{a}}\right] \ ,
\end{eqnarray}
from which the corresponding interaction force is obtained as
\begin{eqnarray}
\label{fortpch} 
{\bf{F}}^{CT}&=&-{\bf\nabla}_{{\bf a}}E^{CT}\nonumber\\
&=& \frac{q}{4\pi a^{3}}\left[\left({\bf{V}}\times{\hat{z}}\right)-3\frac{\left[\left({\bf{V}}\times{\hat{z}}\right)\cdot{\bf{a}}\right]}{a}{\hat{a}}\right] \ .
\end{eqnarray}

From Eq. (\ref{dpplc}), we observe that, within the framework of PQED, the topological source behaves as an electric dipole when interacting with a point-like charge. More specifically, the expression  (\ref{dpplc}) is equivalent to the interaction energy between the charge $q$ and an effective electric dipole, with dipole moment given by ${\bf{d}}=\left({\bf V}\times\hat{z}\right)$ (see Appendix \ref{A}). Interestingly, this result coincides with the interaction energy between a point-like charge and an electric dipole with dipole moment ${\bf{d}}=\left({\bf V}\times\hat{z}\right)$ obtained in standard Maxwell electrodynamics in $(3+1)$ dimensions \cite{FABGH2010}.
In usual Maxwell electrodynamics in $(2+1)$ dimensions, the topological source also behaves as an electric dipole when interacting with a point-like charge \cite{LHCBMCS}.

\begin{figure}[h!]
    \centering
    \includegraphics[width=0.5\textwidth]{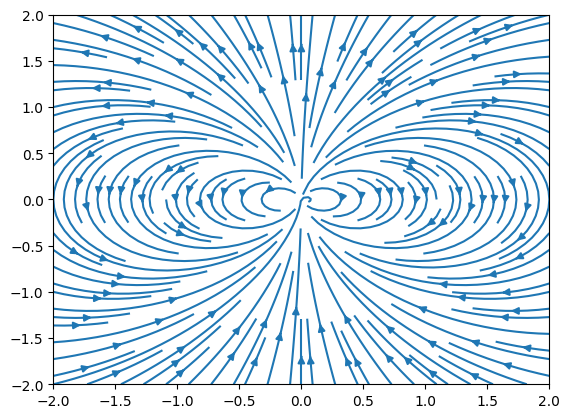}
    \caption{Force lines associated with Eq. (\ref{fortpch})  for the particular configuration, ${\bf V}=V_{x}\hat{x}$. The field lines  are normalized by the factor $1/(V_{x}q)$. The plot clearly exhibits the strongly anisotropic and non-central character of the interaction. The horizontal and vertical axes correspond to the components $a_{x}$ and $a_{y}$, respectively.
}
    \label{fig3}
\end{figure}

In order to illustrate the anisotropic features of the force  (\ref{fortpch}), particularly with respect to the vector ${\bf V}$, we consider the specific configuration ${\bf V}=V_{x}\hat{x}$. In Fig. \ref{fig3}, we display the corresponding force lines obtained for this setup.

\begin{figure}[h!]
    \centering    \includegraphics[width=0.45\textwidth]{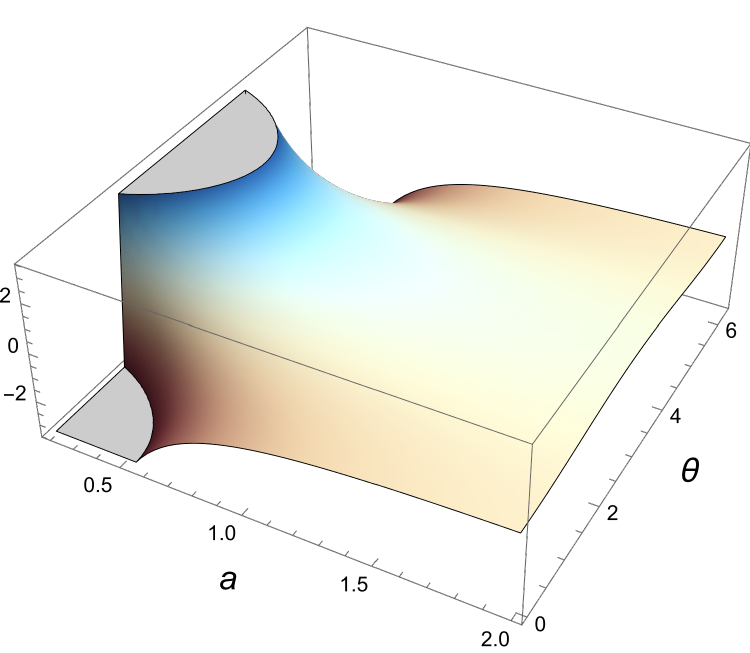}
    \caption{The torque given in Eq. (\ref{torct}), multiplied by $\frac{4\pi}{qV_{x}}$.}
    \label{fig4}
\end{figure}

By keeping the distance $a$ between the point-like charge and the topological source fixed, and for simplicity considering ${\bf V}=V_{x}\hat{x}$, the interaction energy (\ref{dpplc}) gives rise to a nontrivial torque on the system with respect to the vector ${\bf a}$, defined as ${\bf a}=a[\cos(\theta)\hat{x}+\sin(\theta)\hat{y}]$. Under these assumptions, we obtain
\begin{equation}
E^{CT}(\theta)=\frac{q V_{x}}{4\pi a^{2}}\sin(\theta) \ ,
\end{equation}
\begin{eqnarray}
\label{torct}
\tau^{CT}&=&-\frac{\partial}{\partial\theta}E^{CT}(\theta)=-\frac{q V_{x}}{4\pi a^{2}}\cos(\theta) \ .
\end{eqnarray}
In Fig. \ref{fig4}, we have a plot for the torque (\ref{torct}) multiplied by $\frac{4\pi}{qV_{x}}$.

In addition, we observe that the Dirac point does not interact with a point-like charge within the framework of PQED. The same feature is shared by ordinary planar Maxwell electrodynamics. By contrast, such an interaction does arise in the conventional Maxwell--Chern--Simons electrodynamics, where the Dirac point behaves effectively as a point-like charge \cite{LHCBMCS}.

Next, we determine the electromagnetic field configuration generated by a topological source located at the origin, evaluated at an arbitrary point described by the two-dimensional position vector ${\bf r}=\left(r^{1},r^{2}\right)$.

The corresponding field configuration can be obtained from the propagator (\ref{propagator}) according to
\begin{equation}
\label{AMU}
A^{\mu}\left(r\right)=\int d^{3}y \, D^{\mu\nu}\left(r,y\right)J_{\nu}^{T}\left(y\right) \ ,
\end{equation}
where $A^{\mu}=\left(A^{0},A^{1},A^{2}\right)$ and
\begin{equation}
\label{topori}
J^{T}_{\nu}({y})=\epsilon_{\nu}^{\ \alpha\beta}V_{\alpha}\partial_{\beta}
\delta^{2}\left({\bf y}\right) \ .
\end{equation}

By substituting Eqs. (\ref{propagator}) and (\ref{topori}) into Eq.~(\ref{AMU}), and carrying out manipulations analogous to those employed in the previous section, we arrive at
\begin{eqnarray}
\label{AOE}
A^{0}\left({\bf{r}}\right)&=& \frac{1}{2}\left[\left({\bf{V}}\times{\hat{z}}\right)\cdot{\bf{\nabla}}_{\bf{r}}\right]\int\frac{d^{2}{\bf p}}{(2\pi)^{2}}\frac{e^{i{\bf p}\cdot{\bf r}}}{\left({\bf p}^2\right)^{1/2}}\nonumber\\
&=&-\frac{1}{4\pi r^{3}}\left[\left({\bf{V}}\times{\hat{z}}\right)\cdot{\bf{r}}\right] \ ,
\end{eqnarray}
where $r=\mid{\bf{r}}\mid$.

For the spatial components $A^{i}\left({\bf{r}}\right)$ with $i=1,2$, we obtain
\begin{eqnarray}
\label{AKE}
A^{i}\left({\bf{r}}\right)&=&\frac{V^{0}}{2}\epsilon^{i0k}\frac{\partial}{\partial r^{k}}\int\frac{d^{2}{\bf p}}{(2\pi)^{2}}\frac{e^{i{\bf p}\cdot{\bf r}}}{\left({\bf p}^2\right)^{1/2}} \ ,
\end{eqnarray}
which leads us to the following result,
\begin{eqnarray}
\label{ARE}   
{\bf{A}}\left({\bf{r}}\right)=\frac{V^{0}}{4\pi r^{3}}\left({\bf{r}}\times{\hat{z}}\right) \ .
\end{eqnarray}

The electric field can be calculated from the Eq. (\ref{AOE}) as follows, 
\begin{eqnarray}
\label{EA0}    
{\bf{E}}=-{\bf{\nabla}}_{\bf{r}}A^{0}\left({\bf{r}}\right)=\frac{1}{4\pi r^{3}}\left[\left({\bf{V}}\times{\hat{z}}\right)-3\frac{\left[\left({\bf{V}}\times{\hat{z}}\right)\cdot{\bf{r}}\right]}{r}{\hat{r}}\right] \ ,
\end{eqnarray}
where ${\hat{r}}$ is an unit vector pointing in the direction of the vector {\bf{r}}.

The magnetic field can be obtained from Eq.~(\ref{ARE}) as follows,
\begin{eqnarray}
\label{Bbfa}   
B=\epsilon^{0ik}\frac{\partial A^{i}}{\partial r^{k}}=-\frac{V^{0}}{4\pi r^{3}} \ .
\end{eqnarray}

From Eqs. (\ref{EA0}) and (\ref{Bbfa}), we observe that the electric field generated by the topological source is intrinsically anisotropic, since it explicitly depends on the spatial vector ${\bf V}$. In contrast, the corresponding magnetic field is isotropic, depending exclusively on the temporal component $V^{0}$ of the four-vector $V^{\mu}$. 

Furthermore, the magnetic field given in Eq. (\ref{Bbfa}) constitutes an exclusive effect of PQED, since in standard planar Maxwell electrodynamics the topological source does not generate a magnetic field \cite{LHCBMCS}. This result highlights the nontrivial electromagnetic structure induced by the nonlocal character of the gauge sector in PQED.

For the electromagnetic field configuration generated by a Dirac point located at the origin,
\begin{eqnarray}
\label{dpchargeCE}
J^{\nu}_{(D)}\left({ y}\right)=-2\pi i\Phi\int\frac{d^{3}p}{\left(2\pi\right)^{3}}
\delta\left(p^{0}\right)\epsilon^{0\nu\alpha}p_{\alpha}
e^{-i p\cdot y} \ ,
\end{eqnarray}
it is sufficient to consider the previous results obtained for the topological source under the substitutions ${\bf U}={\bf V}=0$ and $V^{0}=-\Phi$. In this limit, Eq. (\ref{EA0}) shows that the Dirac point does not generate an electric field. However, from Eq. (\ref{Bbfa}), we find that the Dirac point produces a magnetic field given by
\begin{eqnarray}
\label{BbfaD}   
B=\frac{\Phi}{4\pi r^{3}} \ ,
\end{eqnarray}
which constitutes an exclusive effect of PQED, since this physical phenomenon has no counterpart in ordinary planar Maxwell electrodynamics \cite{LHCBMCS}.


In summary, in the present letter we investigated the interaction between stationary point-like external sources within the framework of PQED. Besides the usual point-like electric charges, we considered nontrivial localized sources, namely Dirac points and generalized topological sources, and analyzed the corresponding interaction energies, forces, torques, and electromagnetic field configurations generated by these sources.

The generalized topological source considered in the present work may admit interesting interpretations in the context of two-dimensional condensed matter systems, particularly within effective field-theoretical descriptions of graphene and other Dirac materials. In these systems, anisotropic phenomena frequently emerge from lattice deformations, strain-induced gauge fields, disclinations, dislocations, and other topological crystalline defects \cite{Vozmediano2010,Amorim2016,Cortijo2007,CastroNeto2009}. Such mechanisms locally break rotational symmetry and generate direction-dependent electronic and mechanical responses.

From this perspective, the spatial vector sector of the topological source may be interpreted as an effective anisotropic background encoding directional defects or emergent pseudogauge structures in the planar medium. In particular, the intrinsically anisotropic and non-central interactions obtained in this work resemble the directional couplings induced by strain-generated pseudovector gauge fields in graphene \cite{Vozmediano2010,Guinea2010}. Moreover, the emergence of torques and preferred directions in the interaction suggests that generalized topological sources may provide an effective description of anisotropic localized excitations in two-dimensional Dirac systems.
At the same time, the generalized topological source also admits a natural electromagnetic interpretation. As shown in the appendices, its purely spatial sector is equivalent to a pair of point-like electric dipoles, reproducing exactly the corresponding interaction in conventional Maxwell electrodynamics in $(3+1)$ dimensions. Likewise, its purely temporal sector  describes two point-like magnetic dipoles, whose interaction is also identical to that of Maxwell electrodynamics in $(3+1)$ dimensions for magnetic moments oriented perpendicular to the plane. Remarkably, while the electric dipole interpretation also emerges in ordinary planar Maxwell electrodynamics, the magnetic sector has no counterpart in that theory.

Another interesting aspect concerns the limiting case in which the spatial sector of the topological source vanishes, reducing the interaction to the isotropic couplings characteristic of Dirac-point and magnetic dipole. This behavior suggests that the generalized topological source interpolates between isotropic topological excitations and genuinely anisotropic configurations, potentially providing an effective framework for the investigation of defect-mediated directional effects in planar condensed matter systems.

The study of topological sources within the framework of PQED reveals that this planar gauge theory not only modifies the interaction between field sources so that the corresponding power laws agree with those of Maxwell electrodynamics in $(3+1)$ dimensions, but also predicts genuinely new phenomena with no analogue in Maxwell electrodynamics, such as the magnetic field generated by a Dirac point [see Eq.~(\ref{BbfaD})].

Finally, it is worthwhile to comment on possible extensions of the present framework, particularly those involving the inclusion of a Chern--Simons (CS) or pseudo-Chern--Simons term in the gauge sector of PQED. Such a modification would change the tensorial structure of the gauge-field propagator by introducing an antisymmetric contribution and could have important consequences for the interaction between external sources, especially for effects associated with torques. In such a scenario, we anticipate the possibility that the mechanical interaction between the defects may acquire a macroscopic chiral character. It should be stressed, however, that establishing such behavior requires an explicit calculation of the modified PQED propagator in the presence of the CS or pseudo-CS term, followed by the evaluation of the corresponding interaction energies and torques for the external sources considered here. Therefore, the possible emergence of a chiral mechanical response should, at present, be regarded as a physically motivated possibility rather than as a result demonstrated within the model studied in this work. A detailed investigation of this extension is currently in progress and will be presented elsewhere \cite{BorgesBarone}.

\begin{acknowledgments}
L.H.C.B. acknowledges the financial support from Conselho Nacional de Desenvolvimento Científico e Tecnológico (CNPq) and Fundação
de Amparo à Pesquisa do Estado de Minas Gerais-FAPEMIG (Project No. APQ-06536-24). A.A.N. thanks (PROEPD/UDESC) for full support.
\end{acknowledgments}

\appendix  

\section{Point-like electric dipoles}
\label{A}
In this appendix, we investigate the interactions arising from the presence of static point-like electric dipoles within the framework of PQED.

First, we consider a system consisting of two stationary point-like electric dipoles concentrated at fixed points ${\bf {a}}_{1}$ and ${\bf {a}}_{2}$, respectively. The external source for this system reads \cite{FABGH2008,FABGH2010}
\begin{eqnarray}
J_{\mu}^{dd}\left(x\right)=\eta_{\ \mu}^{0}d_{(1)}^{\beta}\partial_{\beta}\delta^{2}\left({\bf {x}}-{\bf {a}}_{1}\right)+\eta_{\ \mu}^{0}d_{(2)}^{\beta}\partial_{\beta}\delta^{2}\left({\bf {x}}-{\bf {a}}_{2}\right)\ ,\label{CDipole}
\end{eqnarray}
with $d_{(1)}^{\beta}=\left(0,{\bf {d}}_{(1)}\right)$ and $d_{(2)}^{\beta}=\left(0,{\bf {d}}_{(2)}\right)$ standing
for fixed and static four vectors, being ${\bf {d}}_{(1)}$ and ${\bf {d}}_{(2)}$ the electric dipole moments. The super index $dd$ denotes that we have two electric dipoles.

By substituting Eqs. (\ref{propagator}) and (\ref{CDipole}) into Eq.(\ref{zxc1}) and proceeding as before, we obtain
\begin{eqnarray}
\label{EnerdEM}
E^{dd}&=&-\frac{1}{2}
\left({\bf{d}}_{(1)}
\cdot{\bf\nabla}_{{\bf a}}\right)\left({\bf{d}}_{(2)}\cdot{\bf\nabla}
_{{\bf a}}\right)\int\frac{d^{2}{\bf p}}{(2\pi)^{2}}\frac{e^{i{\bf p}\cdot{\bf a}}}{\left({\bf p}^2\right)^{1/2}}\
 \ .
\end{eqnarray}
By using Eq. (\ref{int4EM}) and performing the corresponding calculations, we arrive at
\begin{eqnarray}
\label{Enerdd2}
E^{dd}=\frac{1}{4\pi a^{3}}\left[\left({\bf{d}}_{(1)}\cdot{\bf{d}}_{(2)}\right)-3\frac{\left({\bf{d}}_{(1)}
\cdot{\bf{a}}\right)\left({\bf{d}}_{(2)}\cdot{\bf{a}}\right)}{{\bf {a}}^{2}} 
\right] \ .
\end{eqnarray}

We observe that expression (\ref{Enerdd2}) is equivalent to the result obtained in the usual Maxwell electrodynamics in $(3+1)$ dimensions for the interaction energy between two static point-like electric dipoles \cite{FABGH2008,FABGH2010}.

Moreover, formula~(\ref{EnerTEM2}) can be recovered from Eq.~(\ref{Enerdd2}) for the purely spatial configuration
${\bf d}_{(1)}=({\bf V}\times\hat{\bf z})$,
${\bf d}_{(2)}=({\bf U}\times\hat{\bf z})$,
and $V^{0}=U^{0}=0$. This can be seen by noting that, in $(2+1)$ dimensions, the separation vector ${\bf a}$ and the vectors ${\bf U}$ and ${\bf V}$ all lie in the plane perpendicular to $\hat{\bf z}$, so that
\begin{equation}
\label{vetoresplanaresperpz}
{\bf V}\cdot\hat{\bf z}
=
{\bf U}\cdot\hat{\bf z}
=
\hat{\bf a}\cdot\hat{\bf z}
=
0.
\end{equation}
The corresponding electric dipole moments are defined as
\begin{equation}
{\bf d}_{(1)}={\bf V}\times\hat{\bf z},
\qquad
{\bf d}_{(2)}={\bf U}\times\hat{\bf z}.
\end{equation}
So, the energy (\ref{Enerdd2}) reads
\begin{equation}
\label{apEddintermediario1}
E^{\mathrm{dd}}
=
\frac{1}{4\pi a^{3}}
\left\{
\left({\bf V}\times\hat{\bf z}\right)
\cdot
\left({\bf U}\times\hat{\bf z}\right)
-
3
\left[
\left({\bf V}\times\hat{\bf z}\right)\cdot\hat{\bf a}
\right]
\left[
\left({\bf U}\times\hat{\bf z}\right)\cdot\hat{\bf a}
\right]
\right\}.
\end{equation}

Now, we use Lagrange’s identity in order to write
\begin{eqnarray}
\left({\bf V}\times\hat{\bf z}\right)
\cdot
\left({\bf U}\times\hat{\bf z}\right)
=
\left({\bf V}\cdot{\bf U}\right)
\left(\hat{\bf z}\cdot\hat{\bf z}\right)
-
\left({\bf V}\cdot\hat{\bf z}\right)
\left(\hat{\bf z}\cdot{\bf U}\right)={\bf V}\cdot{\bf U},
\end{eqnarray}
where we used Eq.~(\ref{vetoresplanaresperpz}) and the fact that $\hat{\bf z}$ is a unit vector.

From the triple product formula we have
\begin{eqnarray}
\left({\bf V}\times\hat{\bf z}\right)\cdot\hat{\bf a}
=
{\bf V}\cdot\left(\hat{\bf z}\times\hat{\bf a}\right),
\noindent\\
\left({\bf U}\times\hat{\bf z}\right)\cdot\hat{\bf a}
=
{\bf U}\cdot\left(\hat{\bf z}\times\hat{\bf a}\right).
\end{eqnarray}

Collecting terms, Eq.~(\ref{apEddintermediario1}) reads
\begin{equation}
\label{Eddintermediario2}
E_{\mathrm{dd}}
=
\frac{1}{4\pi a^{3}}
\left\{
{\bf V}\cdot{\bf U}
-
3
\left[
{\bf V}\cdot\left(\hat{\bf z}\times\hat{\bf a}\right)
\right]
\left[
{\bf U}\cdot\left(\hat{\bf z}\times\hat{\bf a}\right)
\right]
\right\}.
\end{equation}

Taking into account that the two vectors $\hat{\bf a}$ and $\hat{\bf z}\times\hat{\bf a}$ are mutually orthogonal unit vectors and therefore constitute an orthonormal basis in the plane perpendicular to $\hat{\bf z}$, we can make the decompositions
\begin{eqnarray}
{\bf V}
=
\left({\bf V}\cdot\hat{\bf a}\right)\hat{\bf a}
+
\left[
{\bf V}\cdot\left(\hat{\bf z}\times\hat{\bf a}\right)
\right]
\left(\hat{\bf z}\times\hat{\bf a}\right),
\noindent\\
{\bf U}
=
\left({\bf U}\cdot\hat{\bf a}\right)\hat{\bf a}
+
\left[
{\bf U}\cdot\left(\hat{\bf z}\times\hat{\bf a}\right)
\right]
\left(\hat{\bf z}\times\hat{\bf a}\right),
\end{eqnarray}
and write the scalar product 
\begin{equation}
{\bf V}\cdot{\bf U}
=
\left({\bf V}\cdot\hat{\bf a}\right)
\left({\bf U}\cdot\hat{\bf a}\right)
+
\left[
{\bf V}\cdot\left(\hat{\bf z}\times\hat{\bf a}\right)
\right]
\left[
{\bf U}\cdot\left(\hat{\bf z}\times\hat{\bf a}\right)
\right].
\end{equation}
Hence,
\begin{equation}
\label{apEddintermediario3}
\left[
{\bf V}\cdot\left(\hat{\bf z}\times\hat{\bf a}\right)
\right]
\left[
{\bf U}\cdot\left(\hat{\bf z}\times\hat{\bf a}\right)
\right]
=
{\bf V}\cdot{\bf U}
-
\left({\bf V}\cdot\hat{\bf a}\right)
\left({\bf U}\cdot\hat{\bf a}\right).
\end{equation}

Substituting Eq.~(\ref{apEddintermediario3}) into Eq.~(\ref{apEddintermediario3}) yields
\begin{equation}
\begin{split}
E_{\mathrm{dd}}
&=
\frac{1}{4\pi a^{3}}
\left\{
{\bf V}\cdot{\bf U}
-
3
\left[
{\bf V}\cdot{\bf U}
-
\left({\bf V}\cdot\hat{\bf a}\right)
\left({\bf U}\cdot\hat{\bf a}\right)
\right]
\right\}
\\
&=
\frac{1}{4\pi a^{3}}
\left[
-2{\bf V}\cdot{\bf U}
+
3
\left({\bf V}\cdot\hat{\bf a}\right)
\left({\bf U}\cdot\hat{\bf a}\right)
\right].
\end{split}
\end{equation}
What is exactly expression (\ref{EnerTEM2}) in the case $U^{0}=V^{0}=0$.

In the next situation, we consider the interaction energy between a point-like charge and  an electric dipole. This system is represented by the following external source \cite{FABGH2010}:
\begin{eqnarray}
J_{\mu}^{Cd}\left(x\right)=q\eta_{\ \mu}^{0}\delta^{2}\left({\bf {x}}-{\bf {a}}_{1}\right)+\eta_{\ \mu}^{0}d^{\beta}\partial_{\beta}\delta^{2}\left({\bf {x}}-{\bf {a}}_{2}\right)\ ,\label{CDipoleC} 
\end{eqnarray}
where the superscript $Cd$ indicates that the system is composed of a charge and an electric dipole.

By substituting Eqs. (\ref{propagator}) and (\ref{CDipoleC}) into Eq. (\ref{zxc1}) and performing the corresponding manipulations, we arrive at the following result:
\begin{eqnarray}
\label{Cdintera}   
E^{Cd}&=&\frac{q}{2}\left({\bf{d}}
\cdot{\bf\nabla}_{{\bf a}}\right)\int\frac{d^{2}{\bf p}}{(2\pi)^{2}}\frac{e^{i{\bf p}\cdot{\bf a}}}{\left({\bf p}^2\right)^{1/2}}\nonumber\\
&=&-\frac{q}{4\pi a^{3}}\left({\bf{d}}
\cdot{\bf a}\right) \ ,
\end{eqnarray}
which is equivalent to the result obtained in the usual Maxwell electrodynamics in $(3+1)$ dimensions \cite{FABGH2010}. 
For the purely spatial configuration, with ${\bf d}={(\bf V}\times\hat{\bf z})$, one recovers the interaction energy given by Eq. (\ref{dpplc}).

\section{Point-like magnetic dipoles}
\label{AppMagneticDipole}
In this appendix we compute the interaction energy between two point-like magnetic dipoles within the framework of PQED.

In $(2+1)$ dimensions, the magnetic field is a pseudoscalar rather than a vector. Consequently, the magnetic moment of a point-like source is also described by a pseudoscalar quantity. In the embedding $(3+1)$-dimensional picture, this magnetic moment is interpreted as being oriented perpendicular to the plane.

Let us consider two static point-like magnetic dipoles with magnetic moments $m_{1}$ and $m_{2}$ located at the positions
${\bf a}_{1}$ and ${\bf a}_{2}$, respectively. The corresponding conserved external current is given by
\cite{HelBorgesBarone}
\begin{equation}
J_{mm}^\mu(x)
=
m_{1}\,
\epsilon^{0\mu\nu}
\partial_\nu
\delta^{2}
({\bf x}-{\bf a}_{1}) + m_{2}\,
\epsilon^{0\mu\nu}
\partial_\nu
\delta^{2}
({\bf x}-{\bf a}_{2}) \ ,
\label{CurrentMagneticDipole}
\end{equation}
where the subscript $mm$ denotes a system composed of two magnetic dipoles.

Substituting Eqs. (\ref{propagator}) and (\ref{CurrentMagneticDipole}) into Eq. (\ref{zxc1}) and proceeding along the same lines as in the previous derivations, we arrive at
\begin{equation}
E^{mm}
=
-\frac{m_{1}m_{2}}{2}
\int\frac{d^{2}{\bf p}}{(2\pi)^{2}}{\left({\bf p}^2\right)^{1/2}e^{i{\bf p}\cdot{\bf a}}} \ .
\label{EnergyMomentum2}
\end{equation}

Finally, making use of the result (\ref{int4EMMM})
the interaction energy becomes
\begin{equation}
E^{mm}
=
\frac{m_{1}m_{2}}
{4\pi a^{3}}.
\label{MagneticDipoleEnergy}
\end{equation}
This expression exactly reproduces the interaction energy
between two magnetic dipoles oriented perpendicular to the plane in Maxwell electrodynamics in $(3+1)$ dimensions. For the particular choice $m_{1}=V^{0}$ and $m_{2}=U^{0}$, together with ${\bf V}={\bf U}={ 0}$, one recovers the interaction energy given by Eq. (\ref{EnerTEM2}).

It is important to emphasize that, in ordinary planar Maxwell
electrodynamics, the interaction energy between two magnetic dipoles
described by Eq.~(\ref{CurrentMagneticDipole}) is given by
\begin{eqnarray}
E_{\rm Maxwell}^{mm}
&=&
m_{1}m_{2}
\int\frac{d^{2}{\bf p}}{(2\pi)^{2}}
e^{i{\bf p}\cdot{\bf a}}
\nonumber\\
&=&
m_{1}m_{2}\,
\delta^{2}\left({\bf a}\right),
\end{eqnarray}
where $\delta^{2}({\bf a})$ denotes the two-dimensional Dirac delta
function. Therefore, for spatially separated magnetic dipoles,
${\bf a}\neq{\bf 0}$, the interaction energy vanishes,
\begin{equation}
E_{\rm Maxwell}^{mm}=0,
\qquad {\bf a}\neq{\bf 0}.
\end{equation}
Thus, in ordinary planar Maxwell electrodynamics, two point-like magnetic
dipoles interact only through a contact term. Consequently, the nonvanishing
long-range interaction energy between two spatially separated magnetic
dipoles obtained in PQED has no counterpart in ordinary planar Maxwell Electrodynamics.


\end{document}